\documentclass[twocolumn,showpacs,preprintnumbers,amsmath,amssymb,superscriptaddress]{revtex4}
\usepackage{amsmath, amsfonts, amssymb, mathrsfs}
\usepackage{mathptmx}
\DeclareMathAlphabet{\mathcal}{OMS}{cmsy}{m}{n}
\usepackage{graphicx}
\usepackage{fancybox}
\usepackage{bm}
\usepackage[dvipsnames]{color}
\usepackage{url}
\usepackage[colorlinks, linkcolor=blue, citecolor=black, urlcolor=blue]{hyperref}

\begin{document}


\title{Stabilized Interfacial Ferromagnetism and Enhanced Magnetoelectric Properties of Ultrathin FeRh Films Capped with Heavy Transition Metal Ta}




\author{Guohui Zheng}
\email[E-mail me at:]{ghzheng@gzu.edu.cn}
\affiliation{School of Physics, Guizhou University, Guiyang, China}

\author{Nicholas Kioussis}
\email[E-mail me at: ]{nick.kioussis@csun.edu}
 \affiliation{Department of Physics, California State University, Northridge, CA 91330-8268, USA}


\date{\today}

\begin{abstract}
Thin FeRh film was extensively studied recently,
and an emergent substrate- and capping-dependent interfacial ferromagnetism (FM) was
widely observed in experiments.
However, the voltage modulation of this interfacial ferromagnetism is barely studied,
which would have profound applications in antiferromagnetic (AFM) FeRh-based
magnetoelectric-random access memory (MeRAM).
Using \textit{ab initio} techniques,
we comparatively study the interfacial ferromagnetic properties
and magnetoelectric responses of ultrathin FeRh films
capped by heavy transition metal Ta.
We find that Ta capping reverses the phase stability of ultrathin FeRh film below 1.5 nm
and gigantically stabilizes the ferromagnetic phase and interfacial ferromagnetism.
Besides, small magnetic moment of 2.2 $\mu_B$ for neighboring Fe atoms,
regardless of magnetic configurations and film thickness,
is induced by Ta capping.
Compared with FeRh/MgO bilayers,
magnetic anisotropies
(in-plane for AFM, perpendicular for FM and interfacial-FM reconstructed trilayers)
and magnetoelectric responses of these trilayers are enhanced.
Furthermore, the VCMA behavior in FM phase is changed from $\vee$-shaped to linear.
These findings demonstrate the manipulation of magnetic ordering of FeRh films
with heavy metal capping and electric field
and can promote the application of FeRh alloy
in magnetic memory and antiferromagnetic spintronics.
\end{abstract}


\maketitle


\section{Introduction}\label{Intro}
Magnetic random access memory (MRAM) utilizing nanomagnetic heterostructures
has witnessed a shifting landscape of memory devices~\cite{Chappert07,Bhatti17}.
However, commonly-used writing schemes
with spin-transfer torque effect~\cite{Ikeda10,Miron11} 
is plagued by power-consumption,
which is about $100$ fJ per switch
and two orders higher than conventional CMOS techniques~\cite{Wang13}.
MeRAM (\textbf{M}agneto\textbf{e}lectric \textbf{RAM}) technology
which harnesses the magnetoelectric effect
offers an alternative solution
with ultralow power consumption, highly scalability, and non-volatility
~\cite{Spaldin19,Fusil14}.
In magnetoelectric multiferroic materials,
an electric field can control the magnetism
due to coupled ferromagnetism and ferroelectricity
~\cite{Spaldin19}. 
Two parameters are pivotal to applications of MeRAM:
perpendicular magnetic anisotropy (PMA)
which ensures the stability of information bits against thermal fluctuations,
and voltage-controlled magnetic anisotropy (VCMA) efficiency
$\beta=\Delta MA/E_{ext}$,
where $\Delta MA$ is change of MA under efficient external electric field $E_{ext}$.
To achieve low switching energy below 1 fJ per bit and low write voltage and write error rate,
a large PMA and a VCMA efficiency $\beta> 200$ fJ/(Vm) is required~\cite{Wang13,Manipatruni18}.
In widely studied heavy-metal/ferromagnet/insulator (HM/FM/IN) heterostructures,
HM-dependent PMA and $\beta$ ranging to few hundreds of fJ/(Vm)
are experimentally observed for Fe- and FeCo-based nanojunctions
~\cite{Endo10,Wang12,Rajanikanth13,Nozaki16}.
Recently, a record-high VCMA coefficient exceeding 1000 fJ/(Vm)
for Ir/FeCo/MgO heterostructure was experimentally reported~\cite{Kato18}
and theoretically corroborated by \textit{ab initio} electronic calculations~\cite{Kwon19}.

On the other hand, antiferromagnetic (AFM) materials
with staggered magnetic ordering and ultrafast switching~\cite{Kuhn17}
provide new possibilities for spintronic and MeRAM devices~\cite{Park11,Kosub17,Yan19}. 
Among various AFM materials, bcc-B2 ordered (CsCl-type) FeRh alloys
are extensively studied
due to its intriguing first-order phase transition from AFM to FM order around 370 K,
accompanied with volume expansion of $\sim$ 1\%
and release of large specific heat~\cite{Shirane63,Lewis16}.
In the room-temperature G-AFM phase (abbreviated as AFM later),
Fe atoms have moments around $\pm 3.1$ $\mu_B$ and Rh atoms are non-spin polarized;
while in the high temperature FM phase,
Fe atoms have moment of $3.0$ $\mu_B$ and Rh atoms develop moments of $1.0$ $\mu_B$.
This unique property is harnessed to fabricate
thermal-assisted magnetic memory~\cite{Fullerton03,Marti14}
and thermal cooling devices~\cite{Thiele2011}.
Recently, intensive research interests are focusing on
manipulations of spin order of epitaxially grown ultrathin FeRh films
on MgO~\cite{Bordel12}
and multiferroic~\cite{Cherifi14,ZQLiu2016,Lee15,Amirov19} substrates.
For the FeRh films as-grown,
spin reorientation across the AFM-FM phase transition was observed~\cite{Bordel12},
and (strain-mediated) magnetoelectric coupling
and electric-field-control of the magnetic phase transition were achieved ~\cite{Cherifi14,ZQLiu2016,Lee15,Amirov19}.
Besides the bulk phase transition properties,
interfacial magnetism in
films~\cite{Baldasseroni14,Fan10,Suzuki09}
and nanocrystals~\cite{Hillion13,Loving13}
is also in the spotlight of FeRh study.
In our previous paper~\cite{Zheng17},
we investigated manipulation of the magnetism of
ultrathin FeRh/MgO bilayers
by purely electric field means
(rather than E-field induced strain),
and found large VCMA efficiency
ranging from 200 to 360 fJ/(Vm).
Though the effects of capping metals
on the phase stability and interfacial magnetism of ultrathin FeRh films
are extensively studied,
their effects on magnetoelectric response
are barely studied so far,
while they are of fundamental importance to FeRh-based MeRAM devices.
In this work, we choose a typical heavy transition metal Ta
which is widely used in Fe- and FeCo-based heterostructures
and carry out exhaustive {\it ab initio} electronic structure calculations
to systematically address these issues.

\section{Models and computational methods}\label{Model}
\begin{figure}[htb]
\centering
\includegraphics[width=0.48 \textwidth]{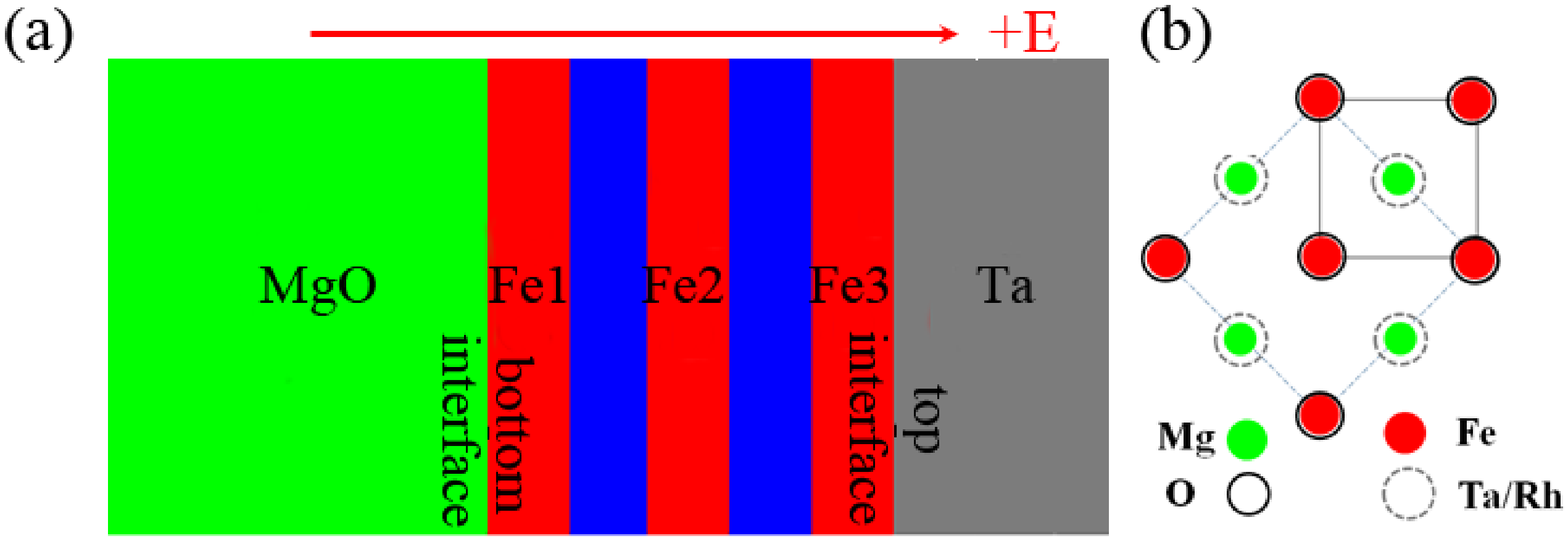}
\includegraphics[width=0.48 \textwidth]{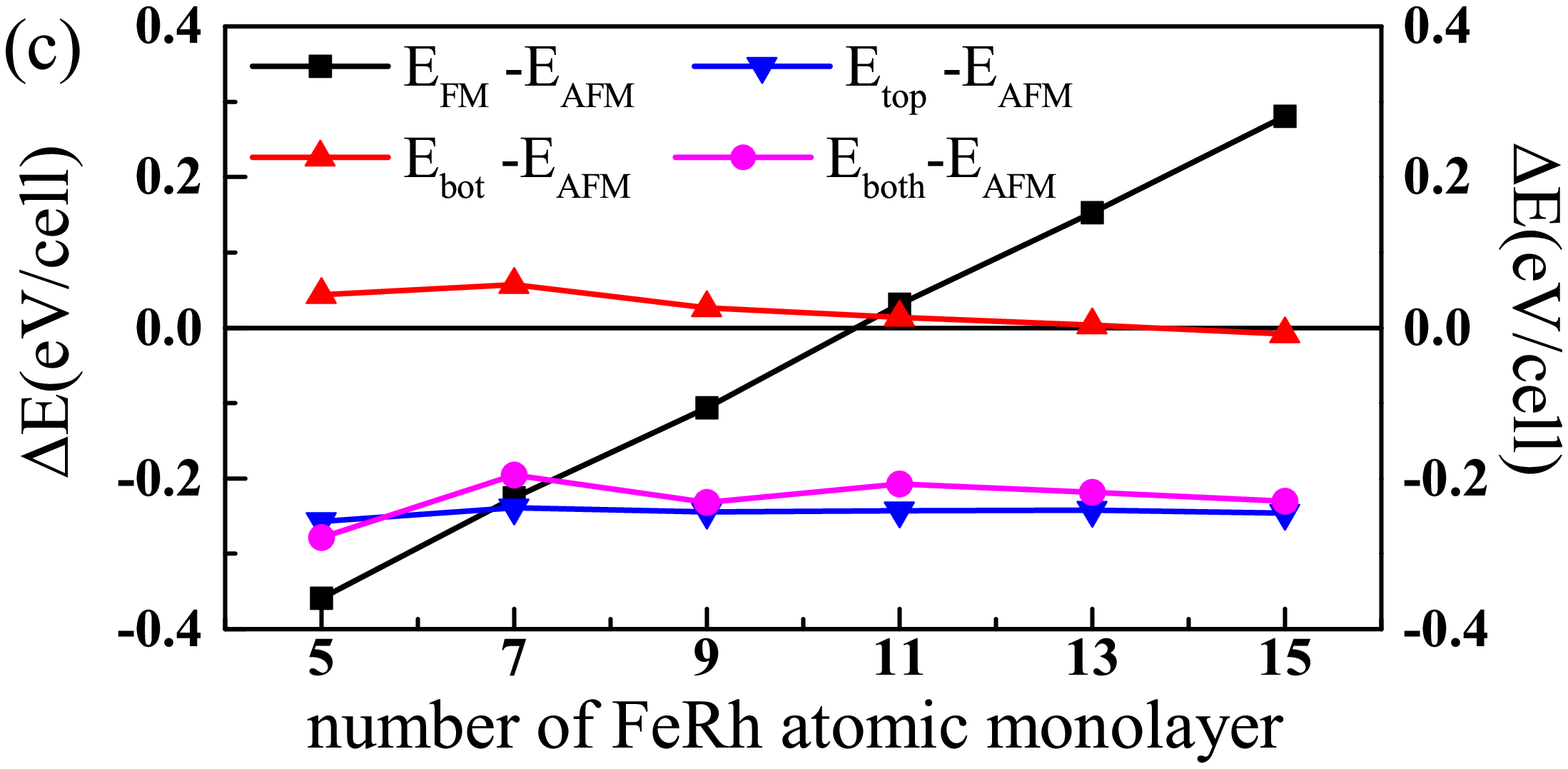}
\includegraphics[width=0.48 \textwidth]{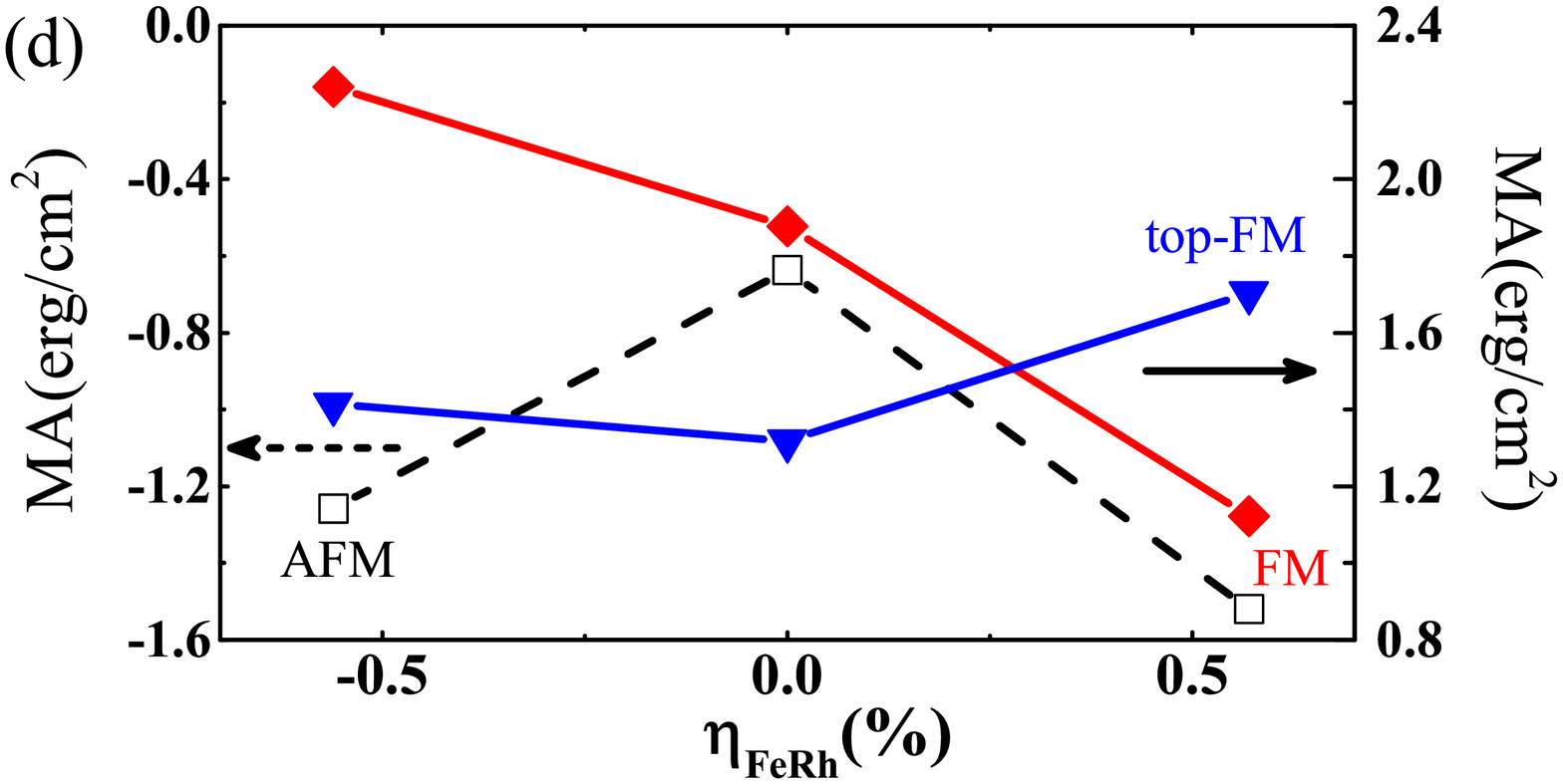}
\caption{(Color online)
(a) Schematic view of Ta/FeRh (5ML)/MgO trilayer structure with two interfaces: top and bottom,
(b) Top view of the two interfaces within the trilayer.
(c) Energetics of various magnetic configurations with lattice constant of 4.212~\AA~
as function of number of FeRh ML,
where top, bot and both indicate the top-, bottom- and both-interface
magnetic-reconstructed phases respectively.
(d) Strain (defined with respect to AFM lattice constant)
modulated zero-field MA of Ta/FeRh (5ML)/MgO trilayer
for AFM (open square, left ordinate) and FM (solid diamond, right ordinate)
and top-FM (solid triangle, right ordinate) phase.
}\label{fig1}
\end{figure}

The ultrathin Ta/FeRh/MgO trilayers are modelled with slab supercell along [001] direction,
and consist of three monolayers (MLs) of bcc Ta on top of five MLs of FeRh
on top of four MLs of rock-salt MgO.
An appropriate vacuum space is included
to avoid spurious interaction between periodic images.
Due to strong spin-orbit coupling of Rh atoms
and low VCMA efficiency of Rh-terminated bilayer structures~\cite{Zheng17},
only Fe-terminated trilayer structures are considered in this paper.
In the Ta/FeRh (top) interface, Ta atoms are placed atop of hollow site of Fe ML;
while in the FeRh/MgO (bottom) interface, the $\mathrm{\langle 110\rangle}$ axis of FeRh
(thus $\sqrt{2} \times \sqrt{2}$ doubling of its slab unit cell)
is aligned with the $\mathrm{\langle 100\rangle}$ axis of MgO
and Fe atoms are placed atop of the O atoms (see Fig.~\ref{fig1}(b)).
We use VASP packages~\cite{Kresse96a, Kresse96b} to implement electronic structure calculations,
and employ PBE functional~\cite{PBE96} to describe the exchange-correlation effect,
and set energy cutoff of the plane-wave expansion of the basis functions to 500 eV.
The dipole layer method~\cite{Makov95,Neugebauer92}
is used to introduce the electric field and to correct the dipole moment in FeRh slab.

The optimized bulk lattice constant $a$,
for the AFM and FM FeRh phases are 2.995~\AA \ and 3.012~\AA, respectively,
in good agreement with experiment~\cite{Shirane63}.
In the slab structure, the lattice mismatch between MgO substrate (4.212~\AA) and FeRh
introduces compressive strain for the FM and AFM FeRh phases,
which have large tuning effects on the VCMA behavior~\cite{Zheng17,Ong15}.
We define the strain with respect to equilibrium lattice constant of AFM bulk phase,
so the three considered supercell lattice constants
(4.212~\AA, 4.236 (=$\sqrt{2} \times 2.995$)~\AA,
4.260 (=$\sqrt{2} \times 3.012$)~\AA)
correspond to strain of -0.57\% (compressive), 0\%, and 0.57\% (tensile) respectively.
For ultrathin trilayers with and without electric field,
using $15 \times 15 \times 1$ $k$-point mesh,
atomic positions along [001] direction and the magnetic and electronic degrees of freedom
for each in-plane lattice constant are fully relaxed
until the Hellmann-Feynman force acting on each ion is less than 0.01 eV/\AA.
For each relaxed structure,
two separate scalar relativistic spin-orbit-coupling (SOC) calculations
with $31 \times 31 \times 1$ $k$-point mesh
and with magnetization axis along the [100] and [001] directions respectively are implemented,
the MA per unit area is then determined from total energy difference $MA = [E^{[100]} - E^{[001]}]/A$,
where $A$ is the in-plane area of the unit cell of slab supercell.

\begin{figure*}[htb]
\centering
\includegraphics[width=0.98 \textwidth]{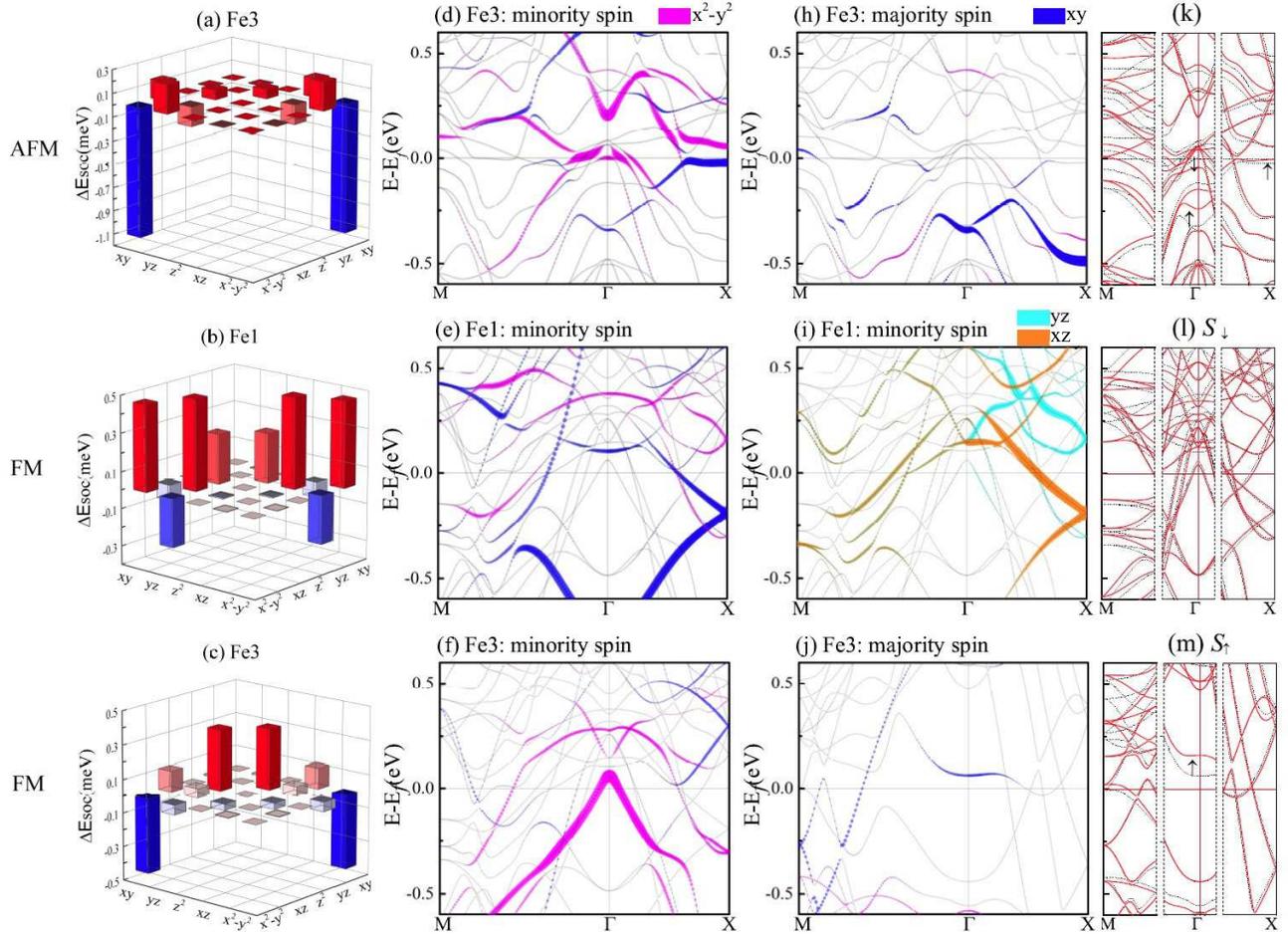}
\caption{(Color online)
(a)-(c) $d$-orbital matrix decomposed $\Delta E_{soc}$ from most-contributed atoms
under $\eta_{FeRh}$ = -0.57\%,
(d)-(j) Energy- and $\mathbf{k}$-resolved distribution
of the $d$-orbital character along symmetry direction,
(k)-(n) Band-shift when the strain changes from
$\eta_{FeRh}$ = -0.57\% (black dash) to $\eta_{FeRh}$ = 0.0\% (red solid),
arrows indicate large band-shifts to determine the MA variation.
}\label{fig2}
\end{figure*}

\section{Results and Discussions}\label{Results}

In ultrathin FeRh films down to nanoscale,
stabilized ferromagnetism induced by finite-sized structural relaxation
was experimentally observed~\cite{Suzuki09,Hillion13,Loving13}.
Therefore, it is fundamentally important to understand
relative stability of various magnetic phases of FeRh thin film under capping
in order to achieve efficient manipulation of its spin order.
Accordingly, we calculate the energetics of various FeRh phases
(AFM, FM, and three interface magnetic-reconstructed phases)
as function of film thickness with MgO lattice constant,
and show the results in Fig.~\ref{fig1}(c).
Energy difference between FM and AFM phases,
$\mathrm{E_{FM}^{total} - E_{AFM}^{total}}$,
increases linearly from negative to positive
with transition occurring at 11 ML of FeRh ($\sim$1.5 nm);
while energy difference between AFM and top-interface magnetic-reconstructed phase
(denoted as top-FM) remains almost constant,
with the top-FM being 250 meV/cell lower than the AFM.
For the bottom- and both-interface magnetic-reconstructed phases
(denoted as bot-FM and both-FM respectively),
their energies are higher than top-FM.
Overall, the FM phase is most stable for thin Ta/FeRh/MgO trilayers less than 6 ML FeRh
and top-FM is the most stable one for thicker films.
These results are in consistent with experimental observations~\cite{Baldasseroni14}.
Hereinafter, we focus on the behaviors of AFM, FM and top-FM phases.
Recalling that in FeRh/MgO bilayer
the AFM phase is always more stable than the FM~\cite{Zheng17,Jekal2015},
we conclude that Ta capping stabilizes the FM gigantically
and even reverse the phase stability for ultrathin films below 1.5 nm,
and the thinner the FeRh film the larger the stabilizing effect.
Therefore, manipulation of the AFM-FM phase transition temperature
of ultrathin FeRh films would be achieved
by growth of Ta capping or FeRh/Ta superlattice.
It's worth noting that
Fe atom near Ta capping develops a small magnetic moment of $\sim 2.1$ $\mu_B$,
regardless of film thickness and magnetic configuration,
which is 1.0 $\mu_B$ lower than Fe atoms in other sites and
in strong contrast with free-standing FeRh film and FeRh/MgO bilayer
where moments change little~\cite{Zheng17}.
Moreover, the Ta atom in FM and top-FM phases
acquires a small moment of $\sim 0.3$ $\mu_B$.
These moment changes can be attributed to
intra-atom charge transfer between majority and minority spin of interfacial Fe,
and inter-atom charge transfer between Fe and Ta atoms~\cite{Odkhuu18}.


Given the interfacial nature of MA in ultrathin films~\cite{Ikeda10,Ong15,Dieny17,Maruyama09},
Ta-induced moment changes would have large effect on MA behaviors of Ta/FeRh/MgO trilayers.
In Fig.~\ref{fig1}(d),
we show MAs of trilayers with 5 MLs of FeRh under different strains.
For AFM phase,
Ta capping not only destroys the linear dependence on strain
for FeRh/MgO bilayer~\cite{Zheng17},
but also enhances its in-plane MA;
while for FM and top-FM phases,
they display enhanced PMA and similar dependence on strain.
The spin reorientation transition from in-plane to
out-of-plane direction across AFM to FM phase
is agreeable with recent experiment~\cite{Bordel12}.

According to Daalderop~\cite{Daalderop94},
MA mainly comes from two distinct contributions:
(i) lowering of the total energy
induced by the splitting of partially occupied degenerate eigenstates
by spin-orbit coupling,
(ii) coupling of eigenstates with energies above and below the Fermi energy
through the spin-orbit interaction.
Through model calculations,
Skomski concluded that the first term gives 50\% to total MA for iron series elements
and nearly 100 \% for rare-earths~\cite{Skomski11}.
Within second-order perturbation theory,
the second contribution has the following form~\cite{Wang93}:
\begin{equation}\label{eq1}
    MA = \xi^{2}\sum_{ou,\sigma\sigma'}(-1)^{\sigma\sigma'}
    \frac{
     {\vert\langle\Psi^{\sigma}_{o}\vert \hat{L}_{z}\vert\Psi^{\sigma'}_{u}\rangle\vert}^2-
     {\vert\langle\Psi^{\sigma}_{o}\vert \hat{L}_{x}\vert\Psi^{\sigma'}_{u}\rangle\vert}^2
     } {E^{\sigma}_{u}-E^{\sigma'}_{o}}
\end{equation}
where $\Psi^{\sigma}_{o}$($E^{\sigma}_{o}$) and $\Psi^{\sigma}_{u}$($E^{\sigma}_{u}$)
are the one-electron occupied and unoccupied states (energies)
of band index $n$ and wave vector $k$
(omitted for simplicity) for $\sigma$ spin ($\downarrow$ and $\uparrow$),
$\xi$ is the SOC constant,
and $\hat{L}_{x(z)}$ is the x(z) component of the orbital angular momentum operator.
For transitions within same spin channels ($\uparrow\uparrow$ and $\downarrow\downarrow$),
$L_z$ gives positive contribution and $L_x$ gives negative contribution,
that is $(-1)^{\uparrow\uparrow}=(-1)^{\downarrow\downarrow}=1$;
while for spin-flip transitions between different spin channels, $(-1)^{\uparrow\downarrow}=(-1)^{\downarrow\uparrow}=-1$.

In order to find out the most-contributed atoms to total MA,
we calculate difference of the atom-decomposed on-site spin-orbit coupling energy
for in- and out-of-plane magnetization orientation,
$\Delta E_{soc}^{i}=E_{soc}^{i,[001]}- E_{soc}^{i,[001]}$
where $E_{soc} =\langle \frac{\hbar^2}{2m^2c^2}\frac{1}{r}\frac{dV(r)}{dr} \hat{L}\cdot \hat{S}\rangle$.
We find that the summation of on-site SOC energy differences from all Fe atoms
quantitatively agrees with the total MA
calculated from the total energy difference.
When adding that from Rh atoms and Ta atoms,
it gives erroneous MA that is two or three times larger in magnitude.
For AFM FeRh films, the Fe3 atom interfaced with Ta capping contributes most;
for FM and top-FM films, the Fe1 atom close to MgO substrate gives largest contribution.

Then we implement second-order perturbation theory in Eq.~\ref{eq1}
to qualitatively understand the origin of MA.
In Fig.~\ref{fig2}(a)-(c)
we plot the $d$-orbital matrix decomposed $\Delta E_{soc}$
of most contributed atoms for these systems,
and in Fig.~\ref{fig2}(d)-(j) we plot the energy- and $\mathbf{k}$-resolved distribution
of the $d$-orbital characters along symmetry paths $\mathrm{M-\Gamma-X}$ for pertinent atoms.
For film in AFM phase,
the inter-spin coupling (manifested by its negative sign)
$\langle  d_{xy}|L_z|d_{x^2-y^2}\rangle$ in Fig.~\ref{fig2}(a)
plays dominating role to determine the sign and magnitude of MA.
For Fe3 atom, appreciable unoccupied $d_{xy}$ orbitals (Fig.~\ref{fig2}(d))
in majority spin channel are developed.
However, for Fe1 and Fe2 atoms with little magnetic moment changes,
these $d_{xy}$ orbitals are absent.
Around $\Gamma$ point,
the newly developed occupied $d_{xy}$ orbitals
strongly couple with unoccupied $d_{x^2-y2^2}$ in minority spin,
and contribute large negative MA here.
Around $\mathrm{X}$ point, the coupling within minority spin channel
between occupied $d_{xy}$ and unoccupied $d_{x^2-y^2}$ through $L_z$
($\langle  d_{xy}|L_z|d_{x^2-y^2}\rangle$) gives positive contribution,
but the inter-spin channel coupling $\langle d_{xy}|L_z|d_{x^2-y^2}\rangle$
gives non-negligible negative MA.
In total, the Ta-induced charge transfer of Fe3 atom
between majority and minority spin channels
induces large negative MA.
When the strain changes from compressive to zero,
we observe significant band shifts (indicated by the arrows in Fig.~\ref{fig2}(k)):
at $\Gamma$ point (I) one band shifts from unoccupied to occupied
and (II) one occupied band move up largely,
at $\mathrm{X}$ point, (III) one band just below Fermi energy moves up even closer.
From the point of view of second-order perturbation theory,
the 1st band shift disables negative inter-spin
$\langle d_{xy}|L_z|d_{x^2-y^2}\rangle$ coupling,
2nd band shift strengthens the negative inter-spin coupling,
and 3rd band shift strengthens the positive intra-spin coupling.
Adding all these up, it explains qualitatively the positive MA variation.

\begin{figure}[b]
\centering
\includegraphics[width=0.27 \textwidth]{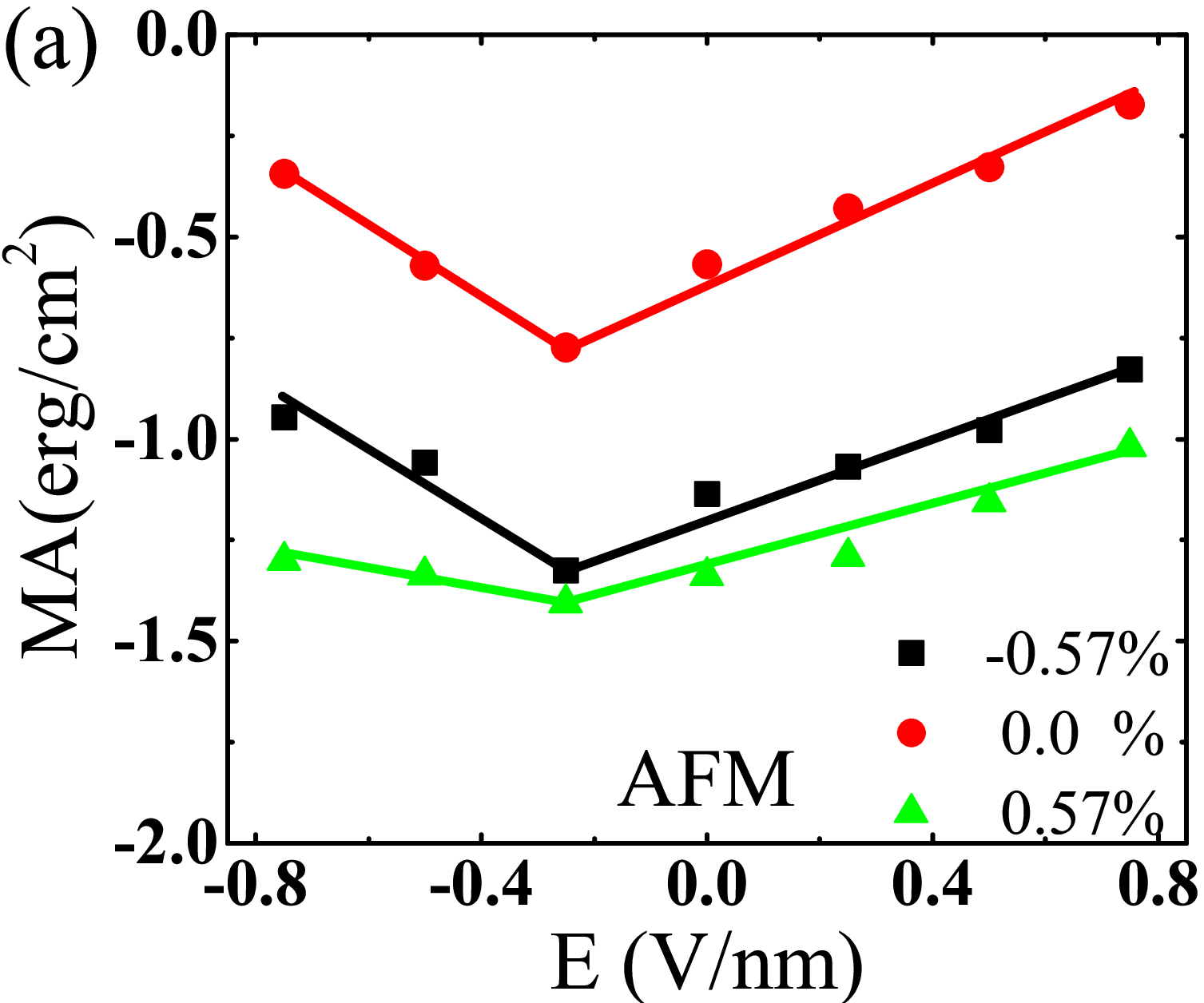}
\includegraphics[width=0.18 \textwidth]{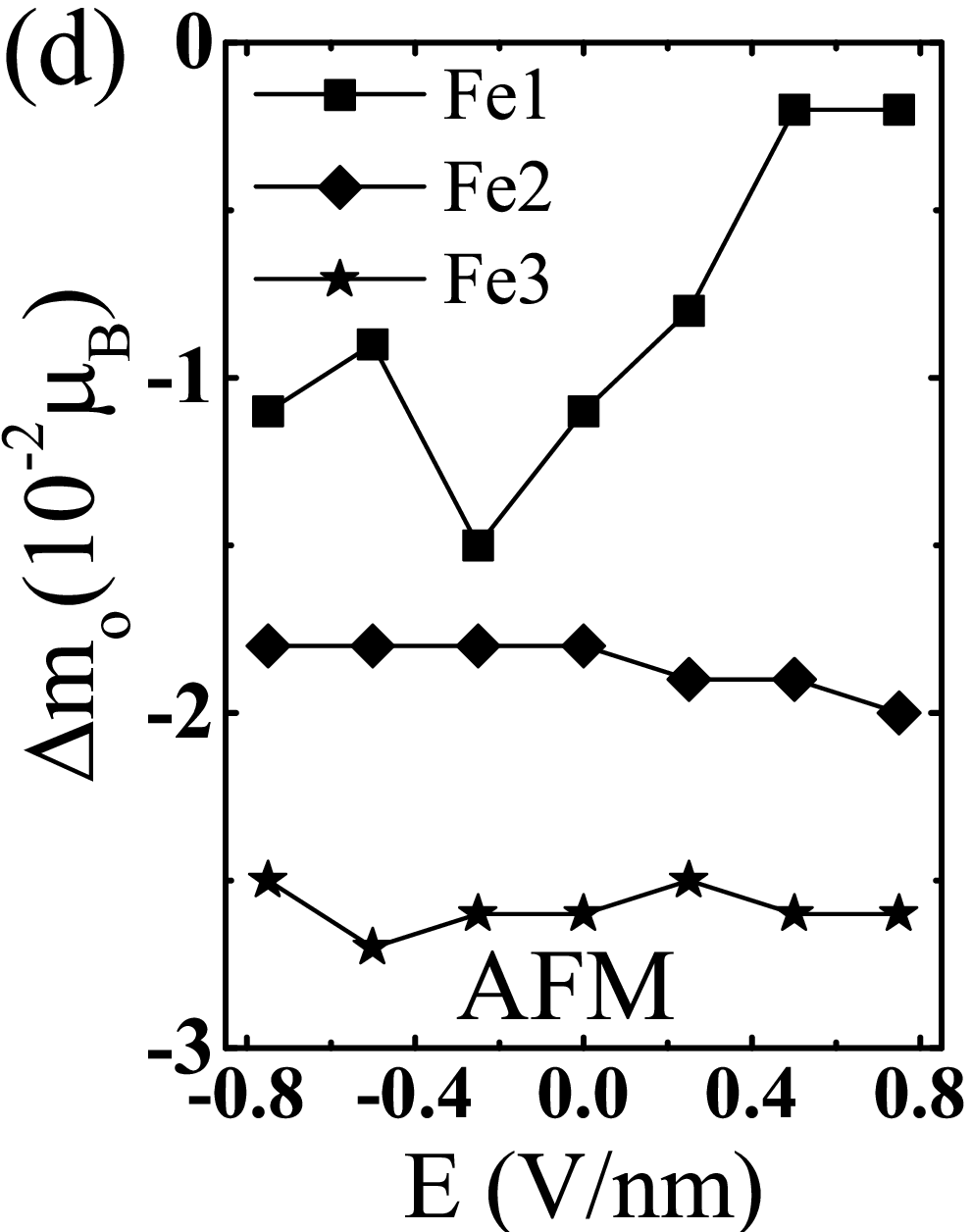}\\
\includegraphics[width=0.27 \textwidth]{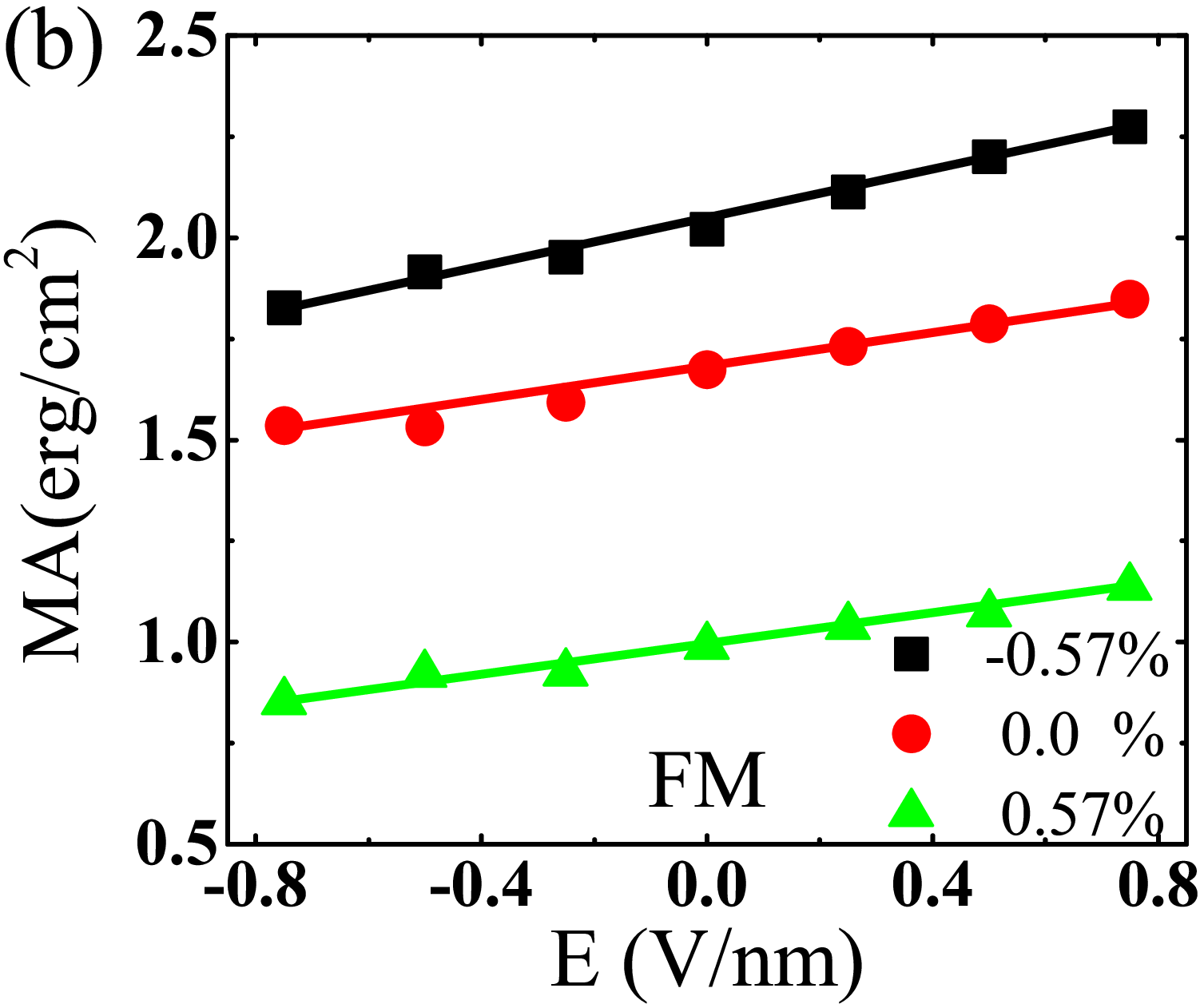}
\includegraphics[width=0.18 \textwidth]{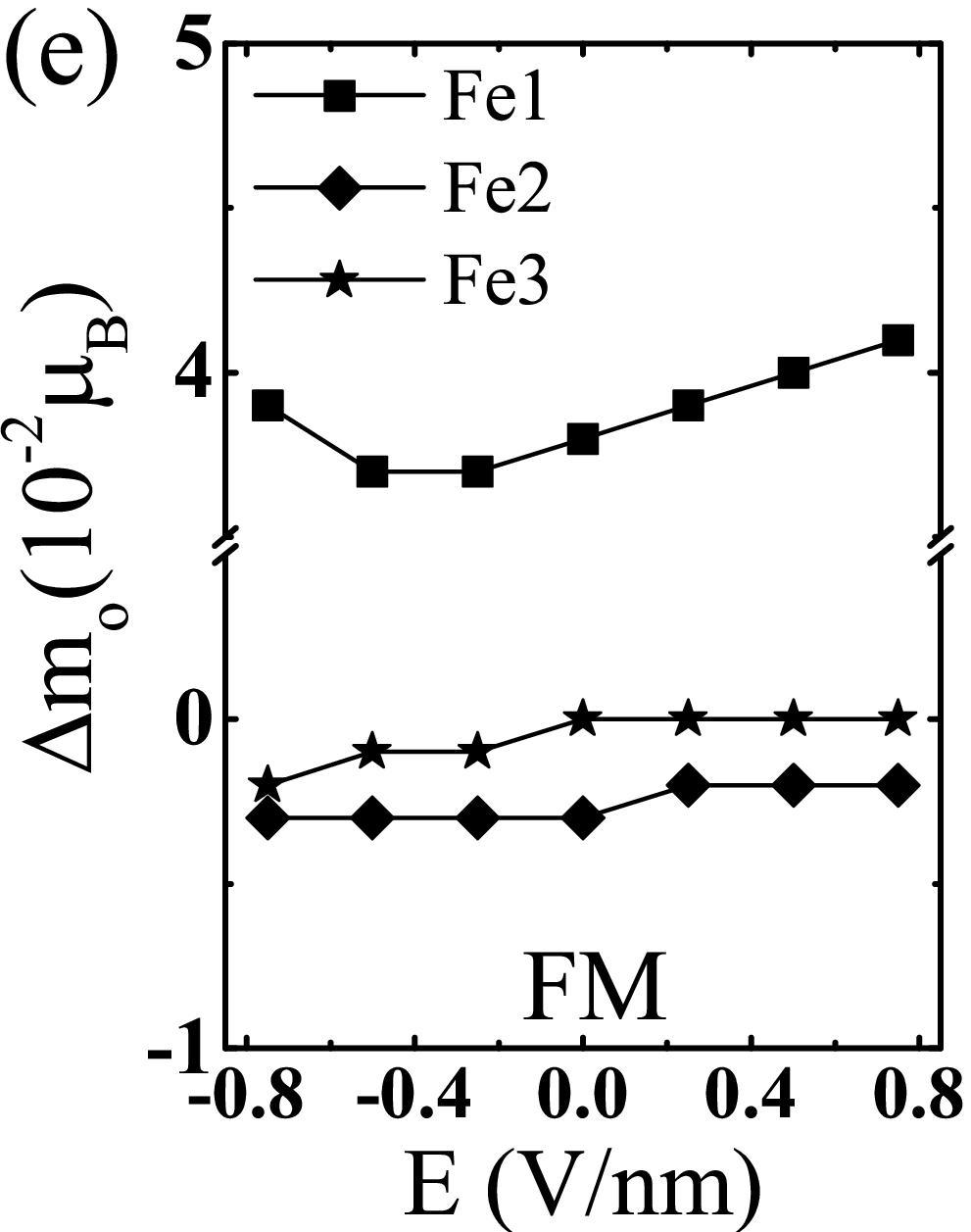}\\
\includegraphics[width=0.27 \textwidth]{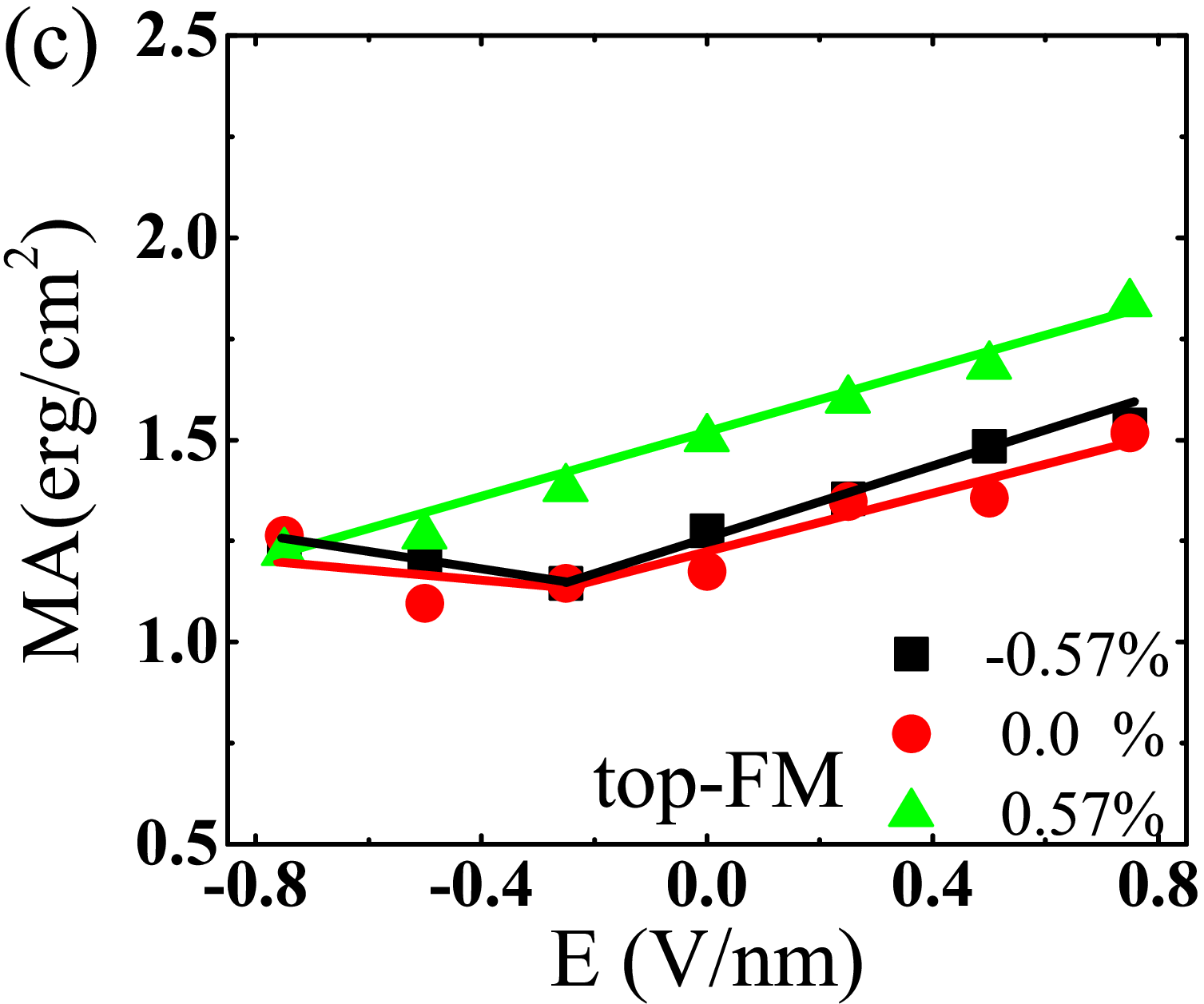}
\includegraphics[width=0.18 \textwidth]{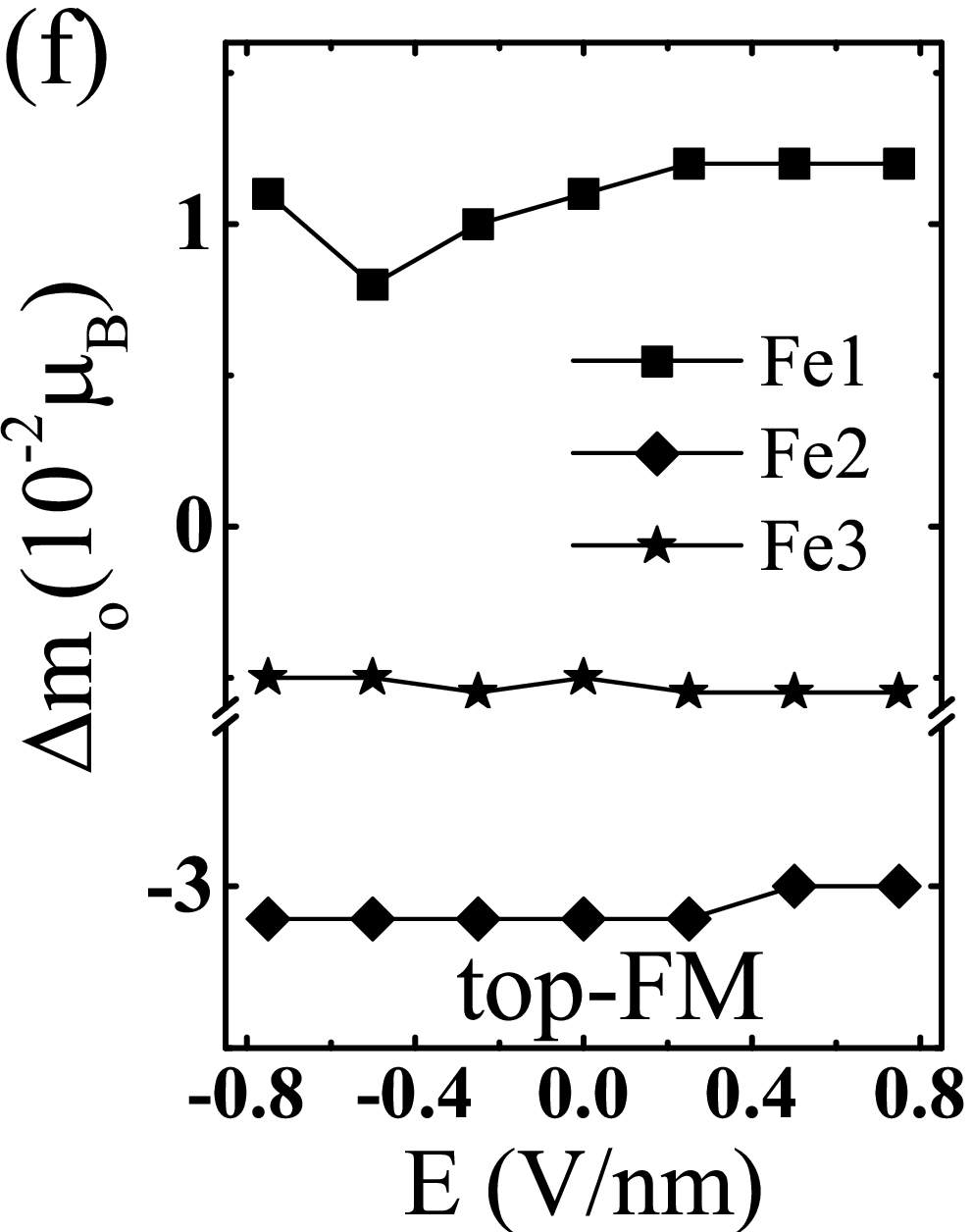}\\
\caption{(Color online) MA versus E-field in MgO
for (a) AFM, (b) FM and (c) top-FM phases under various strain.
The orbital moment difference ($\Delta m_o =m_o^{[001]}-m_o^{[100]}$)
of various Fe atoms versus E-field in MgO
for (d) AFM, (e) FM and (f) top-FM phases under compressive strain of -0.57\%.
}\label{fig3}
\end{figure}

For FM phase,
the scenario that charge transfer between spin channels gives large negative
$\langle d_{xy}|L_z|d_{x^2-y^2}\rangle$ contribution also occurs for Fe3 atom,
as illustrated in Fig.~\ref{fig2}(c), (f) and (j).
However, the large positive MA is by and large determined by the Fe1 atom.
From Fig.~\ref{fig2}(b),(e) and (i),
we find that the positive MA is from the $\langle d_{xy}|L_z|d_{x^2-y^2}\rangle$
and $\langle d_{yz}|L_z|d_{xz}\rangle$ couplings,
which occur around $\mathrm{X}$ and $\mathrm{M}$ point within minority spin channel.
At zero strain (red solid),
the minority spin bands around Fermi energy in Fig.~\ref{fig2}(l) move slightly,
while the majority spin bands move significantly (see Fig.~\ref{fig2}(m)).
In particular,
the unoccupied $d_{xy}$ orbital in majority spin channel around $\Gamma$
are moved up higher,
thus increasing energy spacing in denominator in Eq.~\ref{eq1}
and decreasing negative $\langle d_{xy}|L_z|d_{x^2-y^2}\rangle$ contribution from Fe3 atom.
The strain dependence of FM phase is not as evident as that in AFM phase
from the perspective view of band structure and second-order perturbation analysis
since the minor changes of the dominating minority spin-channels.
For top-FM film, atom-resolved analyses reveal that
contributions from Fe1 and Fe2 atoms to total MA and
their distributions of $d$-orbitals are similar to their counterparts in AFM,
and the behavior of Fe3 atoms are very similar to Fe3 in FM.
It is not surprising since the top-FM configuration is composed of
two AFM layers and one FM layer.


We then explore the VCMA behaviors of various magnetic phases,
the results shown in Fig.~\ref{fig3}(a)-(c) indicate
that both magnetic configuration and strain have large effect on the VCMA behavior.
The VCMA of these trilayer systems in the low-bias regime
are of $\vee$-shaped or linear dependence on the electric field.
VCMA efficiency $\beta$ can be fitted out using linear relation,
$VCMA = \beta E_{MgO} = \beta E_{ext}/\varepsilon$
where $\varepsilon$=10 is the dielectric constant of the MgO
in the range of strain ($\sim \pm$ 0.57\%)~\cite{Ong15},
and $E_{ext}$ is the external E-field.
For AFM trilayer (Fig.~\ref{fig3}(a)),
the magnetization orientation remains in-plane (negative MA)
and the VCMA has a robust asymmetric $\vee$-shape against strain.
The fitting $\beta$ values are
+500 (-860) fJ/(Vm) for positive (negative) E-field for $\eta_{FeRh}$ = -0.57\%,
+630 (-880) fJ/(Vm) for zero strain;
and +370 (-250) fJ/(Vm) for $\eta_{FeRh}$=+0.57\%.
Similar $\vee$-shape E-dependence of MA has been reported
in FM-based trilayers~\cite{Maruyama09,Ong15,Nozaki16}.
The predicted VCMA coefficients are about one to two times larger
than that of AFM FeRh/MgO bilayer~\cite{Zheng17}
and are higher than the critical value of $\sim$ 200 fJ/(Vm)
to achieve ultra-low switching energy
below 1fJ per bit in the next-generation of MeRAMs~\cite{Wang13}.
For the FM trilayer (Fig.~\ref{fig3}(b)),
the MA is linearly dependent on electric field.
The VCMA efficiency $\beta$ values are
190 fJ/(Vm) for $\eta_{FeRh}$ = -0.57\%,
200 fJ/(Vm) for zero strain;
and of 300 fJ/(Vm) for $\eta_{FeRh}$=+0.57\%.
Comparing with FM FeRh/MgO bilayer,
VCMA behaviors are changed from $\vee$-shaped to linear with enhanced coefficients.
Besides, the magnetization of FM trilayer remains out-of-plane (PMA)
regardless of strain for the entire range of E-field,
demonstrating a spin reorientation of the Fe-terminated FeRh film
across the AFM-FM transition.
For the top interface ferromagnetic-reconstructed (top-FM) trilayer (Fig.~\ref{fig3}(c)),
the VCMA behavior changes from $\vee$-shaped to linear
when strain changes from compressive to tensile.
The fitting VCMA coefficiency $\beta$ values are
+440 (-220) fJ/(Vm) for positive (negative) E-field for $\eta_{FeRh}$ = -0.57\%,
+360 (-120) fJ/(Vm) for zero strain;
and +400 fJ/(Vm) for $\eta_{FeRh}$=+0.57\%.


\begin{figure}[htb]
\centering
\includegraphics[width=0.22 \textwidth]{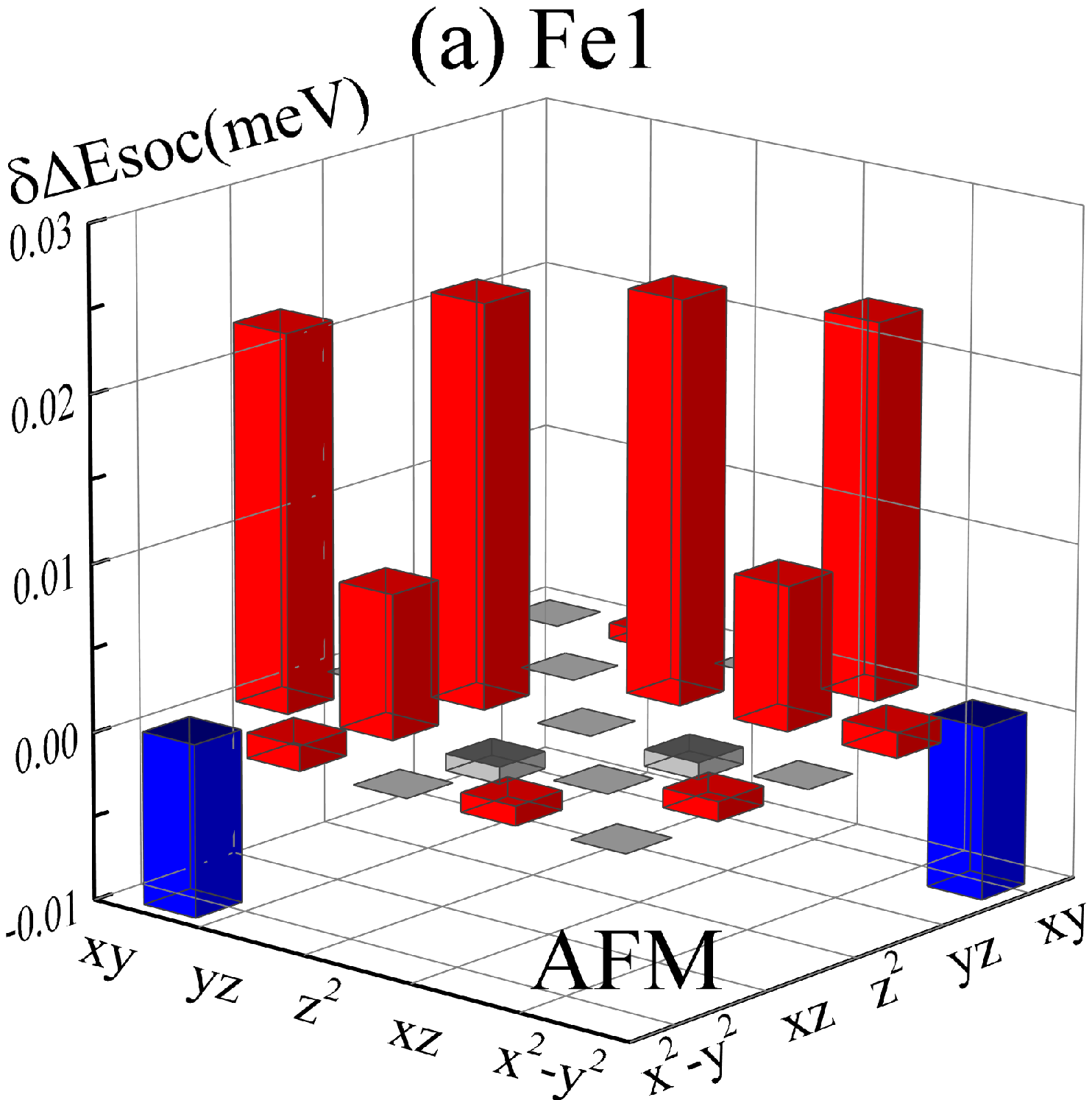}
\includegraphics[width=0.22 \textwidth]{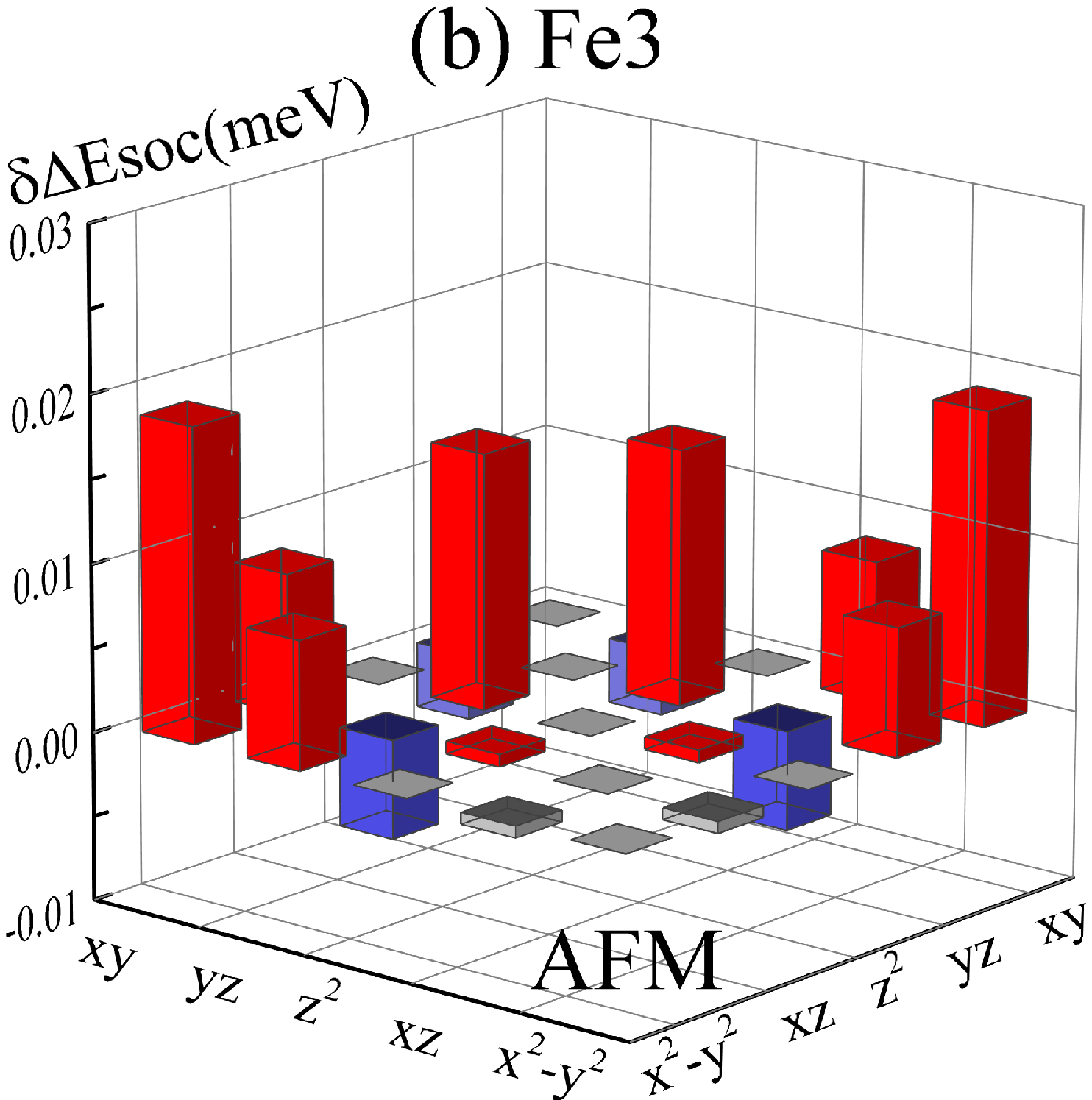}\\
\includegraphics[width=0.22 \textwidth]{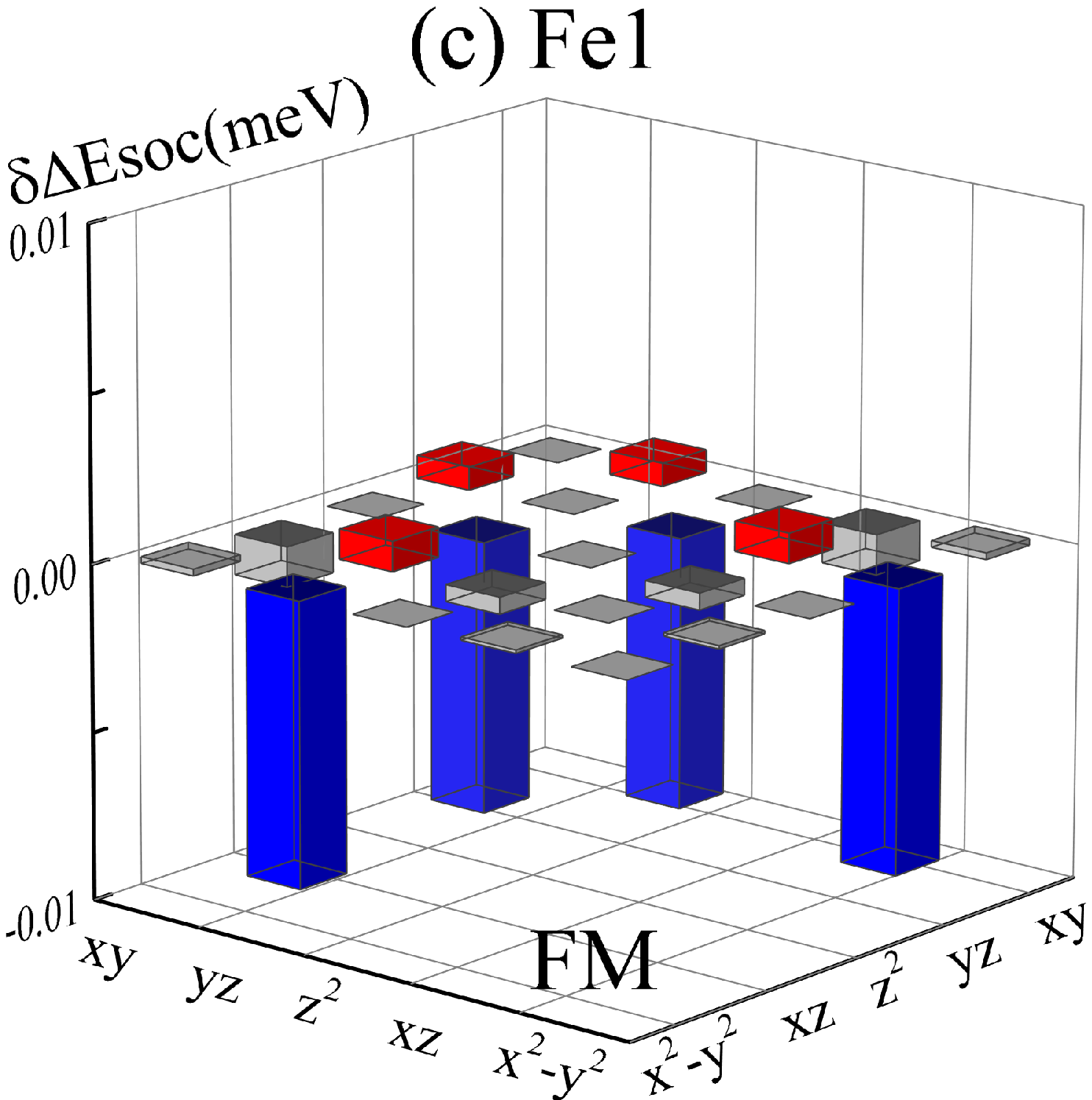}
\includegraphics[width=0.22 \textwidth]{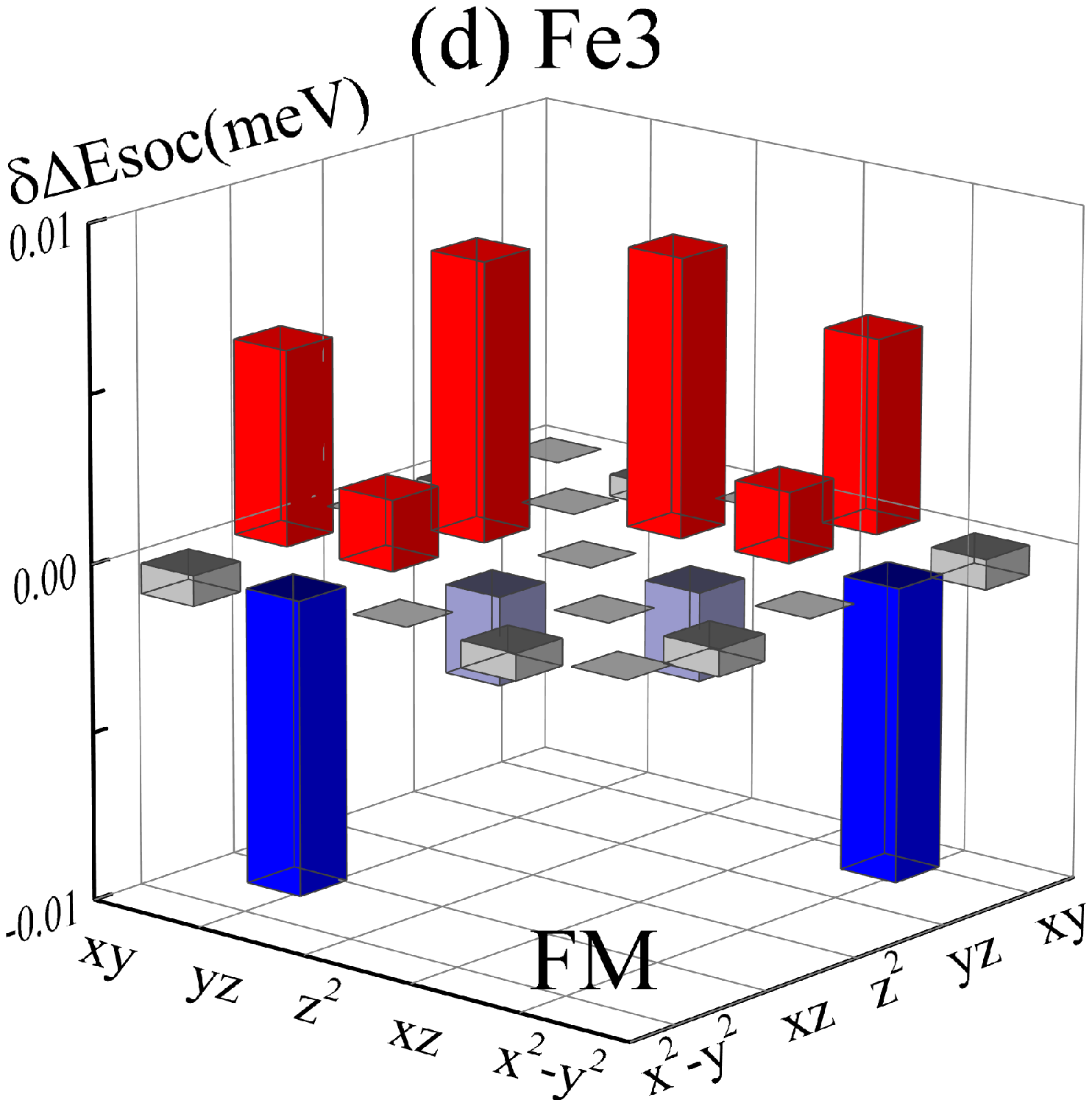}\\
\caption{\label{fig4}
(Color online)
E-field induced change of $\Delta E_{soc}$, that is $\delta(\Delta E_{soc})$,
under -0.75 V/nm for (a) for Fe1 and (b) Fe3 in AFM phase and
(c) for Fe1 and (d) Fe3 in FM phase with compressive strain of 0.57\%}
\end{figure}

In order to probe the mechanism for VCMA,
we show in Fig.~\ref{fig3}(d)-(f)
the orbital moment variation $\Delta m_o =m_o^{[001]}-m_o^{[100]}$
from various Fe atoms with respect to E-field under the compressive strain of 0.57\%,
here $m_o^{[001]([100])}$ is orbital moment along [001]([100]) direction.
The $\Delta m_o$ of Fe1 has strong $\vee$-shaped dependence on external electric field
for these three magnetic configurations,
and others are almost constant.
Besides, correlation between VCMA and $\Delta m_o$ of Fe1 is obvious,
especially for positive E-field,
indicating that electric field affects mostly Fe atom close to MgO substrate.

Neglecting the spin-flip transition between majority and minority spin channels
and using second-order perturbation theory,
Bruno deduced a linear relation (Bruno relation) between MA and $\Delta m_o$,
$MA=\xi \Delta m_o/(4\mu_B)$,
which is highly applicable in ferromagnets
such as Fe, Co and their alloys~\cite{Bruno89}.
We should emphasize that although the AFM nature and non-negligible spin-flip transition
(see Fig.~\ref{fig2}(a),(d) and (h)) in AFM Ta/FeRh/MgO trilayers (Fig.~\ref{fig3}(d)),
the Bruno relation is still applicable.
For FM trilayer in Fig.~\ref{fig3}(e), MA decreases with increasing negative E-field,
and cannot be modeled by Bruno relation
which would rather give increasing MA values due to increasing $\Delta m_o$ of Fe3 atoms.

When E-field is applied across the heterostructure,
electron/hole accumulations at the MgO/Fe1 and Fe3/Ta interfaces
would selectively dope various $d$ orbitals
and change the MA energy through spin-orbit coupling.
In Fig.~\ref{fig4} we show variations of on-site SOC energy difference
($\delta \Delta E_{soc}$) of Fe1 and Fe3 atoms of AFM and FM trilayers
under negative E-field of -0.75 V/nm.
For AFM trilayer, the negative E-field induces large positive
$\langle d_{xy}|L_x|d_{xz}\rangle$ and $\langle d_{z^2}|L_x|d_{yz}\rangle$ variations
both for Fe1 and Fe3 atoms.
For Fe3 atom, there is an additional positive variation from $\langle d_{xy}|L_z|d_{x^2-y^2}\rangle$.
Reminding that previous discussions attribute this matrix element to inter-spin coupling,
we conclude that E-field also affects spin-flip transition.
For FM trilayer, negative E-field of -0.75 V/nm
strengthens the spin-conserved and negative transition
$\langle d_{x^2-y^2}|L_x|d_{yz}\rangle$ (Fig.~\ref{fig2}(b)-(c))
and induces large negative variation for Fe1 and Fe3 atoms.
However, it affects the spin-flip transition
$\langle d_{z^2}|L_x|d_{yz}\rangle$ (Fig.~\ref{fig2}(b)-(c)) differently,
strengthening that of Fe3 atom and weakening that of Fe1 atom.
Adding all these contributions up,
it gives negative MA variation and explains the linear VCMA behavior of FM trilayer.

\section{Conclusions} \label{Conclusion}

In conclusion, using \textit{ab initio} electronic calculations
we comparatively study the properties
of Fe-interfaced Ta/FeRh/MgO nanomagnetic heterojunctions,
and address the effect of Ta capping
on the interfacial magnetism and magnetoelectric coupling of ultrathin FeRh films.
Compared to FeRh/MgO bilayers,
we find that Ta capping in FeRh trilayer
1) gigantically stabilizes the FM phase and interfacial ferromagnetism,
and even reverse the phase stability for ultrathin films below 1.5 nm;
2) induces large charge transfer and smaller magnetic moment of 2.2 $\mu_B$
for neighboring Fe atoms regardless of magnetic configuration,
and enables large spin-flip transition and negative contribution to magnetic anisotropy
when spin-orbit coupling is considered;
3) enhances in-plane MA for trilayer AFM phase
and perpendicular MA for trilayer FM and interfacial FM phase,
and modifies and enhances their magnetoelectric responses upon applied electric field.
More interestingly, we find that
the VCMA behaviors correlate with difference of orbital moment of
Fe atom close to MgO substrate rather than Fe close to Ta capping.
These findings demonstrate the feasibility
to manipulate the magnetic ordering of FeRh film with heavy metal capping and electric field
and have potential applications in AFM FeRh-based MeRAM.

\section{Acknowledgements \label{5}}
G. Z. acknowledges the financial support from Guizhou University (No. (2017)66),
NSF of China (No. 11847049), and Guizhou Province (No. (2018)5871).
N.K acknowledges the financial support from NSF of USA
under Grant No. ERC-Translational Applications of
Nanoscale Multiferroic Systems (TANMS)-1160504.

\end{document}